\documentclass[10pt,prb]{revtex4}
\usepackage{epsfig}

\begin{document}

\title{Diffusion under a flat potential with time dependent sink}
\author{ Diwaker and Aniruddha Chakraborty \\
School of Basic Sciences, Indian Institute of Technology Mandi, Mandi, Himachal-Pradesh 175001, India.}

\begin{abstract}
The Smoluchowski equation for a free particle with a time dependent sink is solved exactly for many special cases. In this method by knowing the probability distribution at the origin $P(0,t)$, one may derive the probability distribution at all positions {\it i.e.,} $P(x,t)$. 
\end{abstract}

\maketitle

\section{Introduction}
\noindent 
The diffusion process of a particle in a potential with the presence of a sink studied through solution of the Smoluchowski equation\cite{diw1} is of keen interest to many scientists in chemical dynamics as it serves  a reference model for a wide variety of dynamical processes. Various attempts has been made to study the diffusion processes with a suitable position of the sink\cite{diw2,diw3} which can find application in diffusion controlled reactions\cite{diw4}. Such a model is used by Wilemski and Fixman\cite{diw5,diw6} to calculate the rate of diffusion controlled reactions as well as cyclization of polymer chain in solutions. Ovchinnikova\cite{diw7}, Zusman\cite{diw8}, Marucs and Nadler\cite{diw9} have used such a model for electron transfer reactions in polar solvents. Marcus\cite{diw10} recently used a diffusive equation with a sink term to develop a theory of uni-molecular reactions in clusters. Pressure influence on isomerization reactions is also explained by Sumi\cite{diw11} by using such type of model.Bagchi, Fleming\cite{diw12,diw13} uses a model of this kind to analyze barrier less electron relaxation in solution.   Exact analytical results for diffusion problems helps in understanding the different parameters like friction and provide an insight to different approximations. As per the authors knowledge, there are no cases where the Smoluchowski equation with a time dependent sink is solved by analytical methods. Even there is no analytical solution available for the simplest possible case {\it i.e.,} for free particle. There are large number of works on diffusion under time independent sink. In recent work from our group we have given an analytical model in which the problem of particle undergoing diffusive motion under the influence of two potentials where the coupling is time independent\cite{diw14}. Most of that has focussed on the time evolution/propagator derivation of the case of one or more Dirac Delta function sinks with constant strength in time. In contrast, this paper presents an work that deals mainly with a Dirac Delta sink whose strength varies with time and we are the first one to consider this effect explicitly. There are two common methods that one may think of for solving this problem. One method is using a path integral method of the Feynman type and the other is using Laplace transforms. We use the latter method though the two methods are closely related. Our method closely follows the method used for solving Schrodinger wave equation.

\section{Formulation of the problem}       

\noindent We would like to solve the Smoluchowski equation with a time-dependent sink as shown in Figure 1.
\begin{figure}
\centering 
{\includegraphics[width=110mm]{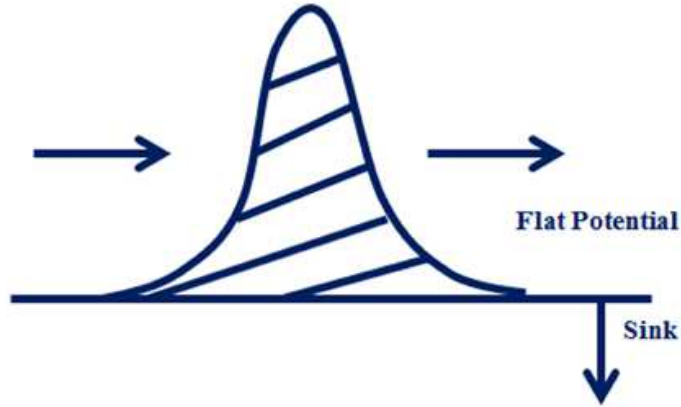}}\caption{
Schematic diagram showing the formulation of our problem}
\end{figure}
\begin{equation}
A\frac{\partial^2 P(x,t)}{\partial x^2}+2k(t)\delta(x)P(x,t)= \frac{\partial P (x,t)}{\partial t}.
\end{equation}
In the above equation $A$ is the diffusion coefficient. It is related to the friction coefficient $\chi$ by $A = \frac{k T}{\xi}$. By integrating Eq. (1) from $x = 0 - \epsilon $ to $x = 0 + \epsilon $, we find that the Eq.(1) reduces to.
\begin{eqnarray}
A \frac{\partial^2 P(x,t)}{\partial x^2}= \frac{\partial P(x,t)}{\partial t} \;\;\;\;\; {\text and} \\ \nonumber \frac{\partial P(0^{+},t)}{\partial x} - \frac{\partial P(0^{-},t)}{\partial x} = 2k(t)P(0,t).
\end{eqnarray}
This is just the case of zero potential with an additional time-dependent boundary condition. If we take the Laplace transform \cite{diw15} of Eq. (2), we get
\begin{equation}
A \frac{\partial^2 {\bar P}(x,s)}{\partial x^2}= s {\bar P}(x,s) - P(x,0),\;\;\; \ {\bar P}(0^{+},s)- {\bar P}(0^{-},s)= 2 L [k(t)\ P(0,t)].
\end{equation}
For solving Eq. (3) we first consider the homogeneous solution in order to satisfy the boundary condition at the origin. The solution is given below
\begin{equation} 
{\bar P}(x,s)= a(s)exp (-\sqrt{s/A}|x|)+\frac{1}{2 A \sqrt{s/A}}\int dx'(-\sqrt{s/A}|x - x'|)P(x',0),
\end{equation}
where $a(s)$ is the integration constant. By applying the boundary condition at the origin to Eq. (4) we find.
\begin{eqnarray}
& \frac{{\bar P}(0^{+},s)}{\partial x}- \frac{{\bar P} (0^{-},s)}{\partial x}  = 2\sqrt{ s/A } a(s) = 2 L\left[k(t) P(0,t)\right]\\ \nonumber
 & \Rightarrow a(s) = \frac{L\left[k(t)P(0,t)\right]}{\sqrt {s/A}}.
\end{eqnarray}
We now set $x=0$ in Eq. (4) to get
\begin{equation} 
{\bar P}(0,s)= \frac{L\left[k(t)P(0,t)\right]}{\sqrt {s/A}}+\frac{1}{2 A \sqrt{s/A}}\int dx'(\sqrt{s/A}|x'|)P(x',0),
\end{equation}
Using the convolution theorem for Laplace transforms \cite{diw16}, we invert Eq. (6) to get
\begin{equation}
P(0,t)-\;\frac 1{\sqrt{\pi/A}}\int_ 0^t\;{dt'}\frac{ k(t) P(0,t')}{\sqrt {t-t'}}= \frac{1}{2 A^2 \sqrt{\pi t/A}}\int_{-\infty}^{\infty}exp\left (A\frac{(x')2}{4t}\right  )P(x',0).
\end{equation}
The Eq.(7) may be used to determine the probability distribution at the origin ({\it i.e.,} $P(0,t)$). Once $P(0,t)$ is known, one may find $P(x,t)$ by inverting Eq.(4) which has the following result:\\

\begin{eqnarray}
\ {P}(x,t)\;=\frac{1}{\sqrt{\pi/A}}\int_ 0^t \;{dt'}\frac{k(t)P(0,t')}{A^2\sqrt {t-t'}}exp\left(A\frac{x^2}{4{(t-t')}}\right)\nonumber \\
+\frac 1{2 A^2\sqrt{\pi t/A}}\int_{-\infty}^{\infty}dx'exp\left(A\frac{(x-x')2}{4t}\right)P(x',0).
\end{eqnarray}
Eq. (7) and Eq.(8) together completely determine the problem. As we can see knowing the probability distribution at the origin {\it i.e.,} $P (0,t)$ is essentially equivalent to knowing the probability distribution everywhere ($P(x,t)$).

\subsection{Derivation of the propagator for time independent sink \it{ i.e.,} $k(t) = k_0$}
\noindent
In some cases it may be advantageous to write down the solution in terms of Propagator as follows
\begin{equation}
P(x,t)=\int_{-\infty}^{\infty}\;dx'G(x,x',t)P(x',0),
\end{equation}
where $G(x,x',t)$ is propagator. To find the propagator, we use Eq.(5) and Eq.(6) to get
\begin{equation}
a(s)=\frac{k_0 \; \bar P(0,s)}{\sqrt{s/A}} = \frac{k_0}{\sqrt{s/A}+  k_0}\frac{1}{2  A \sqrt{s/A} }\int dx'( \sqrt{s/A}|x'|)P(x',0).
\end{equation}
Using Eq. (4) and Eq. (10), we get
\begin{eqnarray} 
{\bar P}(x,s)= \frac{k_0}{\sqrt{s/A}+  k_0} \frac{1}{2  A \sqrt{s/A}}\int dx'exp \left( \sqrt{s/A}(|x|+|x'|)\right)P(x',0)\nonumber \\+\frac{1}{2 A \sqrt{s/A}}\int dx' exp( \sqrt{s/A}|x - x'|)P(x',0),
\end{eqnarray}
so that
\begin{equation}
{\tilde G}(x,x's)=\frac{k_0}{\sqrt{s/A}+  k_0} \frac{1}{2  A \sqrt{s/A}} exp \left( \sqrt{s/A}(|x|+|x'|)\right)+\frac{1}{2 A \sqrt{s/A}}exp( \sqrt{s/A}|x - x'|).
\end{equation}
We now invert Eq. (11) to get
\begin{eqnarray}
G(x,x',t)=\frac{k_0}{2A^2}\;{exp{[k_0(|x|+|x'|)}}+{{k_0}^2 t/A}] \times erfc\left(k_0 \sqrt{ t/A}+\frac{|x|+|x'|}{2\sqrt{ t/A}} \right)\nonumber  \\
 +\frac{1}{2 A^2 \sqrt{\pi t/A}}exp \left( A\frac{(x-x')^2}{4t}\right).
\end{eqnarray}
We can see easily that the same propagator can be found by using Feynman approach. Let us assume that $P(x',0)= \delta(x+a)$, so that $P(x,t)= G(x,-a,t)$ and hence the detailed expression for $P(x,t)$ can be
\begin{eqnarray}
P(x,t)=\frac{k_0}{2A^2}\;{exp{[k_0(|x|+|a|)}}+{{k_0}^2 t/A}] \times erfc\left(k_0 \sqrt{ t/A}+\frac{|x|+|a|}{2\sqrt{ t/A}} \right)\nonumber  \\
 +\frac{1}{2 A^2 \sqrt{\pi t/A}}exp \left(A\frac{(x+a)^2}{4t}\right).
\end{eqnarray}

\subsection{Solution for $k(t)$ linear in time}
\noindent We now consider the case, where $k(t)$ has the following form:
\begin{eqnarray}
k(t)= - \alpha t,\;\;\;t \ge 0. \\ 
\end{eqnarray}
Here $\alpha$ is a real number. We take the initial condition as
\begin {equation}
P(x,0)= \delta (x+a),
\end{equation}
where $a$ and $\alpha$ are real positive numbers. Using Eq.(6)and the identity $L[tf(t)] = - d L[t\times f(t)]/ds$, we get
\begin{equation}
{\tilde P}_s(0,s)= \frac{1}{\alpha}\sqrt{\frac{s}{A}}{\tilde P}(0,s)- \frac{1}{2 A \alpha} exp\left(\sqrt{s/A}|a|\right).
\end{equation}
We can solve this equation under the condition that the Laplace transform vanishes at s=$\infty$ and that it must also be bounded at $\pm i \infty$ .The last condition is required for the inversion integral to exist. So that we get
\begin{eqnarray}
{\tilde P}(0,s)= - \frac{1}{2 A \alpha} exp\left(\frac{1}{\alpha}\sqrt{\frac{s}{A}}\right) \int_{-  \infty}^{s}ds'exp\left(-\frac{1}{\alpha}\sqrt{\frac{s'}{A}}\right) exp\left(\sqrt{\frac{s'}{A}}|a|\right).
\end{eqnarray}
This Laplace domain expression can not be converted to time domain in terms of elementary or special functions. However, it can be done using the integral
\begin{equation}
P(0,t)= \frac{1}{2 \pi i} \int^{i \infty+\beta}_{-i\infty+\beta} ds {\tilde P}(0,s)exp(st).
\end{equation}
In the above expression $\beta$ is a positive number taken to be larger than the right-most pole in the complex plane. So we are unable to do the last step analytically, although numerically it is not a problem.

\subsection{Solution for $k(t)$ inversely proportional to time}

\noindent We choose $k(t)=\alpha/t$, where $\alpha$ is a real positive constant. The Laplace transform of $k(t)$ does not exist. Interestingly, the Laplace transform of $k(t)P(0,t)$ does exist if we choose an initial condition that vanishes at the origin! Our derivation in this subsection is valid under this retriction. We first use the fact,
\begin{equation}
L[\frac{1}{t}f(t),s]=\int^\infty_s ds' L[f(t),s'].
\end{equation}
Now by using Eq. (21) and Eq. (6) we get
\begin{equation}
{\tilde P}(0,s)=\frac{\alpha}{\sqrt{s/A}}\int^\infty_s ds'{\tilde P}(0,s')+\frac{1}{2 A \sqrt{s/A}}\int^\infty_{-\infty} dx' exp (\sqrt{s/A}|x'|)P(x',0).
\end{equation}
Now we define a new function $Q(s)$ as given below
\begin{equation}
Q(s)= \int_{s}^{\infty}ds'{\tilde P}(0,s).
\end{equation}
Now Eq.(22) can be written as
\begin{equation}
\frac{\partial Q(s)}{\partial s}+\frac{\alpha}{\sqrt{s/A}}Q(s)= - \frac{1}{2 A \sqrt{s/A}}\int^\infty_{-\infty} dx' exp (\sqrt{s/A}|x'|)P(x',0).
\end{equation}
On solving Eq. (24) we get,
\begin{equation}
Q(s)=\int^\infty_{-\infty} dx' \frac{exp(\sqrt{s/A}|x'|)}{|x'|+2i\alpha}\psi(x',0)
\end{equation}

By Equation 29 we find,\\

\begin{equation}
\bar{\psi}(0,s)=\frac{1}{2\sqrt{s/A}}\int^\infty_{-\infty} dx'\frac{|x'|exp(\sqrt{s/A}|x'|)}{|x'|+2i\alpha}\psi(x',0)
\end{equation}

We may now invert Equation 30 to find,\\

\begin{equation}
\psi(0,t)=\frac{1}{2\sqrt{i\pi t}}\int^\infty_{-\infty} dx'\frac{|x'|}{|x'|+2i\alpha}\; exp\left(i\frac{(x')^2}{4t}\right)\psi(x',0)
\end{equation}

We now seek $\psi(x,t)$.To find this quantity, we first write,\\

\begin{equation}
\psi(x,t)=\int^\infty_{-\infty} dx' G(x,x',t)\psi(x',0)
\end{equation}

To find G, we use results of Equations (30), (27), (5) and (4) to find,\\

\begin{equation}
\bar{G}(x,x's)=\frac{1}{2\sqrt{is}}\left[-\frac{2i\alpha}{|x'|+2i\alpha}exp(i\sqrt{is}(|x|+|x'|))+exp(i\sqrt{is}(|x-x'|))\right]
\end{equation}

We may now invert this expression to find,\\

\begin{equation}
G(x,x't)=\frac{1}{2\sqrt{i\pi t}}\lfloor-\frac{2 i\alpha}{|x'|+2 i\alpha}\; exp\; \left(i\frac{(x|x|+|x'|)^2}{4t}\right)+exp\left(i\frac{(x-x')^2}{4t}\right)\rfloor
\end{equation}

\subsection{Solution for $k(t)$  with an exponential dependence on time}

\begin{equation}
\ L[exp(at)\;f\;(t),s\;]= L[f\;(t),s\;-a].
\end{equation}
By equation (31) and (6) we find
\begin {equation}
\ {\tilde p}(0,s)=\frac{\sqrt{c_0}}{s-ic^{2}_{0}}\frac{\beta}{\sqrt{s}\;-\;{\sqrt{i c_0}}}{\tilde p}\;(0,s+\alpha)
\end {equation}
To obtain a solution, we could iterate this expression repeatedly to obtain the series solution:\\
\begin {equation}
\ {\tilde p}S(x)=\sum_{n=0}^{\infty}\beta^n a_n(s).
\end {equation}
But since we already Know the form of the serise, we may substitute equation(37)directly into equation(36). After doing this nad solving for the a's by equating terms with like powers of $\beta$ , we find that $a_n$ (s) is given by\\
\begin{center}
\begin{eqnarray}
 \ a_n(s)=\frac{\sqrt{c_0}}{s-ic2_0},\;\;\;\;\;\; n=0,
\end{eqnarray}
\end{center}
\begin{eqnarray}
\ a_n(s)= \frac{(-\sqrt{i)}^{n}{\sqrt{c0}}}{s+n\alpha-ic^{2}_{0}}\prod_{j=0}^{n-1}\frac {1}{\sqrt{s+j\alpha}}-{\sqrt{ic_0}},n>0.
\end{eqnarray}

\section{Conclusions}
In the current work the authors had attempted  to find the exact solution for the Smoluchowski equation with a time dependent delta function sink in many special cases. These special cases include Linear, inversely  and exponential dependence. The case of sinusoidal time dependent sink can be treated in a similar manner as the exponential time dependent sink by representing the sine function as a sum of complex exponentials.

\section{Acknowledgements}
Th authors want to thank IIT Mandi for Professional development fund as well as HTRA scholarship.

\end{document}